\documentclass[letterpaper,twocolumn,showpacs,pra,aps]{revtex4-1}

\usepackage{amsmath}  
\usepackage{amsfonts} 
\usepackage{graphicx} 
\usepackage{amssymb}

\begin{document}

\title{Space charge and screening of a supercritical impurity cluster in monolayer graphene}

\author{Eugene B. Kolomeisky$^{1}$ and Joseph P. Straley$^{2}$}

\affiliation
{$^{1}$Department of Physics, University of Virginia, P. O. Box 400714,
Charlottesville, Virginia 22904-4714, USA\\
$^{2}$Department of Physics and Astronomy, University of Kentucky,
Lexington, Kentucky 40506-0055, USA}

\date{\today}

\begin{abstract}
A Coulomb impurity of charge $Ze$ is known to destabilize the ground state of undoped graphene with respect to creation of screening space charge when $Z$ exceeds the critical value $Z_c = 1/2\alpha$ set by the material's fine structure constant $\alpha$.   Recent experimental advances have made it possible to explore this transition in a controlled manner by tuning $Z$ across the critical point.  Combined with relatively large value of $\alpha$ this makes it possible to study graphene's screening response to a supercritical impurity $Z\alpha\gg1$ when the screening charge is large using a Thomas-Fermi analysis.  The character of screening in this regime is controlled by the dimensionless screening parameter $Z\alpha^{2}$.  Specifically, for a circular impurity cluster in the weak-screening regime $Z\alpha^{2}\ll1$ most of the screening charge is found to reside outside the cluster.  The strong-screening regime $Z\alpha^{2}\gg1$ provides a realization of the Thomson atom:  most of the screening charge is inside the cluster, nearly perfectly neutralizing the source charge with the exception of a transition layer near the cluster's edge where the rest of the space charge is localized.           
          
\end{abstract}

%%\pacs{71.20.Tx, 72.10.Fk, 81.05.ue}

\maketitle

\section{Introduction}

Monolayer graphene is a two-dimensional semimetal whose low-energy elementary excitations obey a pseudo-relativistic dispersion law
\begin{equation}
\label{graphene_dispersion}
\varepsilon(\textbf{p})=\pm v_{F} p
\end{equation}
where $\textbf{p}$ is the two-dimensional momentum vector, $p=|\textbf{p}|$, the upper and lower signs correspond to the conduction and valence bands, respectively, and $v_{F}\approx c/300$ is the Fermi velocity \cite{graphene_review}.  

This paper studies the screening response of undoped pristine graphene to a large external impurity charge $Ze$ (which we may assume to be positive without the loss of generality).  Fundamental interest in this problem can be quickly appreciated by trying to understand the binding properties of the Coulomb field.  Indeed, the ground-state energy of graphene's electron of typical momentum $p$ localized within a spatial scale $r\simeq\hbar/p$ can be estimated by optimizing the energy cost of localization according to Eq.(\ref{graphene_dispersion}) against the potential energy gain in the electric field of charge $Ze$
\begin{equation}
\label{energy_estimate}
\varepsilon(p)\simeq v_{F}p-\frac{Ze^{2}}{\kappa r}\simeq v_{F}p(1-Z\alpha), ~~~\alpha=\frac{e^{2}}{\kappa\hbar v_{F}}\approx\frac{2.5}{\kappa}
\end{equation}
where $\kappa$ is the dielectric constant due to graphene's own electrons and the surrounding environment, and $\alpha$ is the material's fine structure constant \cite{graphene_review}.  Minimizing with respect to the parameter $p$ one can then see that when the dimensionless external charge $Z$ (hereafter simply referred as "charge") is sufficiently small, $Z\alpha\lesssim 1$, the electron is delocalized, $p\propto 1/r=0$, while sufficiently large charge, $Z\alpha\gtrsim1$, leads to an infinitely sharp localization $p\propto1/r=\infty$.  

Essentially the same conclusion applies to the charge state of a point nucleus of charge $Ze$ in vacuum, where now the fine structure constant $\alpha=e^{2}/\hbar c=1/137$ is significantly smaller: the electric field of a highly charged nucleus ( $Z\gtrsim 137$) destabilizes the vacuum  with respect to creation of electron-positron pairs; the positrons escape to infinity leaving behind a screening space charge of electrons \cite{Popov}. Creating a charge exceeding $137 e$ is a serious obstacle to experimental observation of the vacuum instability, but for the suspended graphene the predicted threshold value  $Z_{c}=1/2\alpha\approx \kappa/5$ \cite{Shytov} is experimentally accessible, since $\kappa\approx5$ and thus $\alpha\approx 1/2$ \cite{suspended}), resulting $Z_{c}\approx1$.     

Various aspects of atomic collapse in graphene have been verified in experiments with clusters of charged $Ca$ dimers \cite{collapse}.  More recently a tunable version of the external charge $Z$ was realized \cite{Eva_Andrei}:  a single-atom vacancy in graphene was demonstrated to stably host a local charge and this charge could be gradually built up by applying small voltage pulses with the tip of a scanning tunneling microscope (STM).  The experiment has been performed using a variety of surfaces, specifically, G/G/BN [graphene (G) on graphene on Boron nitride (BN)], G/BN and G/G on $SiO_{2}$, and each sample was independently characterized regarding its value of the parameter $\alpha$ (\ref{graphene_dispersion}).  This allowed to systematically study a range of charges both below and above $Z_{c}=1/2\alpha$, up to $Z\approx1.6/\alpha$ (G/G/BN surface).  When $Z$ exceeds $Z_{c}$ but is still close to it, there are very few screening electrons present and one faces a difficult few-body problem of the binding of graphene's electrons  to a Coulomb center.  A simplification arises in the many-electron case ($Z\gg 1$), when the Thomas-Fermi (TF) method applies \cite{LL3}.  Its main advantages are physical transparency and rigor:  correlation effects vanish to leading order as $Z\rightarrow \infty$ so that a mean-field treatment of electron-electron interactions suffices \cite{Lieb}.  Here we revisit the TF theory for a supercritical $Z\alpha \gg 1$ impurity cluster in graphene.   This regime also appears to be within experimental reach of the approach of Ref. \cite{Eva_Andrei};  it seems plausible that a large charge corresponding to $Z\alpha \gg 1$ can be built up by applying voltage pulses with the STM tip to individual vacancies of a vacancy cluster. 

TF theory in question has been analyzed by several authors \cite{Mele,Katsnelson,Fogler} with conflicting predictions which were attributed to physically different regimes of parameter values \cite{Fogler}.  A comprehensive physical picture of the effect, including analysis of the geometry relevant to most promising experimental setup \cite{Eva_Andrei} is, in our judgement, still lacking.  Our contribution below brings completeness and definiteness to the problem of screening of a $Z\alpha\gg 1$ impurity in graphene.  Our analysis builds on a study due to Migdal \textit{et al.} \cite{MVP} of the QED version of the related problem that was adopted \cite{KSZ} to the case of narrow band-gap semiconductors \cite{Keldysh} and Weyl semimetals \cite{AB}.  Latter systems may be viewed as three-dimensional counterparts of graphene, and below we indeed find that the physics of supercritical impurity screening in graphene resembles that of a Weyl semimetal \cite{KSZ}.  The difference between the problems is largely technical and traceable to two-dimensionality of graphene's electrons interacting according to the three-dimensional Coulomb law.

We hasten to mention that the approach of Refs.\cite{MVP} and \cite{KSZ} has been also employed to solve the problem of the collapse of the electrons to a donor cluster in $SrTiO_{3}$ \cite{BIS}.  

\section{Classical electrostatics}

The total electrostatic potential $\varphi(\textbf{r})$ felt by an electron confined to a plane embedded into three-dimensional space is due to an external potential $\varphi_{ext}(\textbf{r})$ and to the potential caused by the rest of the screening electron cloud of number density $n(\textbf{r})$:
\begin{eqnarray}
\label{sc_potential}
\varphi(\textbf{r})&=&\varphi_{ext}(\textbf{r})-\frac{e}{\kappa}\int\frac{n(\textbf{r}')d^{2}r'}{|\textbf{r}-\textbf{r}'|}
\nonumber\\
&=&\frac{1}{\kappa}\int\frac{\sigma(\textbf{r}')d^{2}r'}{|\textbf{r}-\textbf{r}'|}=\frac{e}{\kappa}\int\frac{[n_{ext}(\textbf{r}')-n(\textbf{r}')]d^{2}r'}{|\textbf{r}-\textbf{r}'|}
\end{eqnarray}
where $\textbf{r}$ is the two-dimensional position vector, $\sigma(\textbf{r})=e[n_{ext}(\textbf{r})-n(\textbf{r})]$ is the surface charge density, and an external number density $n_{ext}(\textbf{r})$ is related to $\varphi_{ext}(\textbf{r})$ as: 
\begin{equation}
\label{ext_potential}
\varphi_{ext}(\textbf{r})=\frac{e}{\kappa}\int\frac{n_{ext}(\textbf{r}')d^{2}r'}{|\textbf{r}-\textbf{r}'|}.
\end{equation}
The integrals are over the plane where the charge resides, but the Coulomb interaction has a three-dimensional form since the fields extend into space.  We note that external charges positioned off the plane also create in-plane $\varphi_{ext}$;  in such cases Eq.(\ref{ext_potential}) defines an equivalent in-plane $n_{ext}$.  Given $\varphi_{ext}$, Eq.(\ref{ext_potential}) can be inverted to provide an explicit expression for $n_{ext}$.  In a compact form this can be given in terms of the Fourier transforms of the potential, $\varphi_{ext}(\textbf{q})$, and the density
\begin{equation}
\label{n_ext_Fourier_solution}
n_{ext}(\textbf{q})=\frac{\kappa}{2\pi e}q\varphi_{ext}(\textbf{q})
\end{equation} 
where $q=|\textbf{q}|$.  For the external potential of charge $Z$ one has $\varphi_{ext}(\textbf{q}\rightarrow 0)=2\pi Ze/\kappa q$ thus implying that $n_{ext}(\textbf{q}\rightarrow 0)=Z$ which is a re-statement that we are dealing with net charge $Z$.  
For example, for a point charge $Z$ that is a distance $a$ away from the plane the external in-plane potential is
\begin{equation}
\label{phi_ext_charge_above_plane}
\varphi_{ext}(\textbf{r})=\frac{Ze}{\kappa(r^{2}+a^{2})^{1/2}}.
\end{equation}  
Then $\varphi_{ext}(\textbf{q})=(2\pi Ze/\kappa q)e^{-qa}$ and following Eq.(\ref{n_ext_Fourier_solution}) one finds $n_{ext}(\textbf{q})=Ze^{-qa}$.  Inverting the Fourier transform gives the real-space density
\begin{equation}
\label{n_ext_charge_above_plane}
n_{ext}(\textbf{r})=\frac{Za}{2\pi(r^{2}+a^{2})^{3/2}}.
\end{equation}  
As another example, let us consider in-plane charge distribution that is Gaussian in the Fourier representation, $n_{ext}(\textbf{q})=Ze^{-q^{2}a^{2}}$.  Then it is also Gaussian in real space:
\begin{equation}
\label{n_ext_Gaussian}
n_{ext}(\textbf{r})=\frac{Z}{4\pi a^{2}}e^{-r^{2}/4a^{2}}.
\end{equation} 
Following Eq.(\ref{n_ext_Fourier_solution}), the external potential is $\varphi_{ext}(\textbf{q})=(2\pi Ze/\kappa q)e^{-q^{2}a^{2}}$ whose inversion \cite{GR} supplies its real-space version 
\begin{equation}
\label{phi_ext_4_Gaussian_n_ext}
\varphi_{ext}(\textbf{r})=\frac{\sqrt{\pi}}{2}\frac{Ze}{\kappa a}e^{-r^{2}/8a^{2}}I_{0}\left (\frac{r^{2}}{8a^{2}}\right )
\end{equation}    
where $I_{0}(y)$ is modified Bessel function of the first kind.  

Eq.(\ref{sc_potential}) makes it clear that the regime of infinitesimally weak screening is
\begin{equation}
\label{zero_screening_limit}
\varphi(\textbf{r})=\varphi_{ext}(\textbf{r}), ~~~n(\textbf{r})=0,
\end{equation}
while the regime of infinitely strong screening is
\begin{equation}
\label{infinite_screening_limit}
\varphi(\textbf{r})=0,~~~n(\textbf{r})=n_{ext}(\textbf{r}).
\end{equation}
For example, for point charge $Z$ positioned a distance $a$ away from ideally conducting (i.e. screening) plane, $n=n_{ext}$ given by Eq.(\ref{n_ext_charge_above_plane}) is the textbook result for the density of induced charge derived by the method of images.  

\begin{figure}
\begin{center}
\includegraphics[width=1\columnwidth]{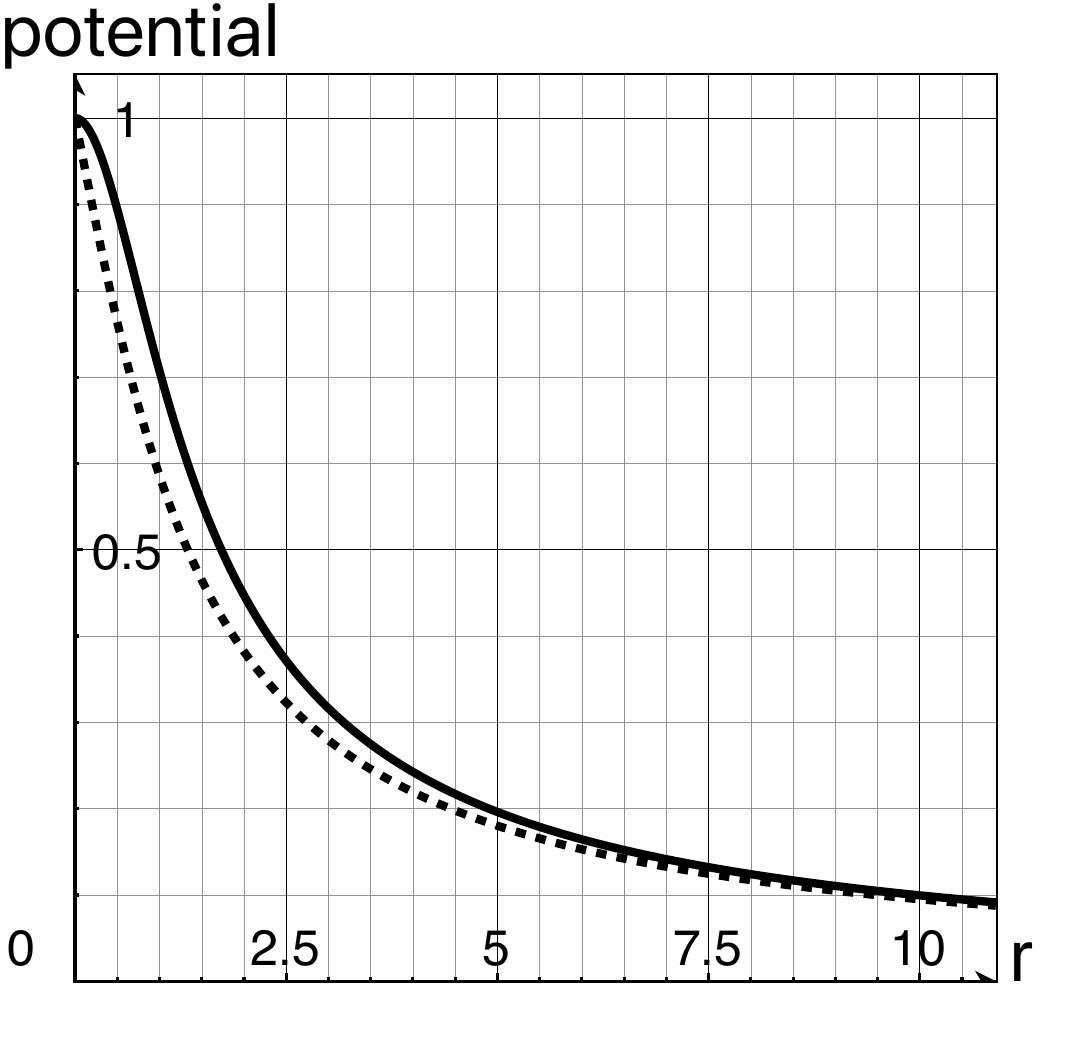}
\caption{In-plane electrostatic potential (in units of $Ze/\kappa a$) versus distance (in units of $a$) due to a point charge outside conductive plane:  exact result (\ref{phi_ext_charge_above_plane}) (bold curve) versus its approximation (\ref{appr}) (dashed curve).}
\label{elliptic}
\end{center}
\end{figure}

When the source has circular symmetry angular integration can be carried out in Eq.(\ref{sc_potential}) with the result \begin{equation}
\label{circular_sc_potential}
\varphi(r)=\frac{4}{\kappa r}\int_{0}^{r}\sigma(r')r'dr'\textbf{K}\left (\frac{r'}{r}\right )
+\frac{4}{\kappa}\int_{r}^{\infty}\sigma(r')dr'\textbf{K}\left (\frac{r}{r'}\right )
\end{equation}
where $\textbf{K}(y)$ is the complete elliptic integral of the first kind and we employed Landen's transformation \cite{GR}.  On the other hand, electrostatic potential of a three-dimensional spherically symmetric charge density distribution $\rho(r)$ can be written as
\begin{equation}
\label{spherical_sc_potential}
\varphi(r)=\frac{4\pi}{\kappa r}\int_{0}^{r}\rho(r')r'^{2}dr'+\frac{4\pi}{\kappa}\int_{r}^{\infty}\rho(r')r'dr'.
\end{equation}
Comparing Eq.(\ref{circular_sc_potential}) with its three-dimensional counterpart (\ref{spherical_sc_potential}) highlights mathematical difference between screening problems in two and three dimensions.  Indeed, multiplying both sides of (\ref{spherical_sc_potential}) by $r$ and twice differentiating the outcome, one recovers the Poisson's equation $\partial^{2}(r\varphi)/r\partial r^{2}=-4\pi \rho/\kappa$.  Similar exact reduction to a differential form is impossible with Eq.(\ref{circular_sc_potential}) thus leaving us with singular integral equation.  Nevertheless a useful simplification of Eq.(\ref{circular_sc_potential}) capturing the physics of the problem is possible and based on the observation that $\textbf{K}(y)$ is nearly constant and slightly exceeds $\pi/2$ over most of its range with the exception of the narrow vicinity of $y=1$ where it has integrable (logarithmic) singularity.  Then approximating $\textbf{K}(y)\rightarrow \textbf{K}(0)=\pi/2$ in the integrands in Eq.(\ref{circular_sc_potential}) reduces the latter to the form 
\begin{equation}
\label{approximate_circular_sc_potential}
\varphi(r)\approx\frac{2\pi}{\kappa r}\int_{0}^{r}\sigma(r')r'dr'
+\frac{2\pi}{\kappa}\int_{r}^{\infty}\sigma(r')dr'
\end{equation}       
which matches the exact result (\ref{circular_sc_potential}) both for $r\rightarrow 0$ and $r\rightarrow \infty$.  To illustrate its accuracy for general $r$, we computed the approximate potential (\ref{approximate_circular_sc_potential}) corresponding to the density (\ref{n_ext_charge_above_plane}) with the result
\begin{equation}
\label{appr}
\varphi_{ext}^{(appr)}(r)=\frac{Ze}{\kappa a}\frac{r+a-\sqrt{r^{2}+a^{2}}}{r}.
\end{equation}
Its comparison with the exact potential (\ref{phi_ext_charge_above_plane}) displayed in Figure \ref{elliptic} shows that Eq.(\ref{approximate_circular_sc_potential}) is a good approximation to Eq.(\ref{circular_sc_potential}) for all $r$.

Multiplying both sides of Eq.(\ref{approximate_circular_sc_potential}) by $r$ and twice differentiating the outcome one obtains equivalent differential equation
\begin{equation}
\label{2dPoisson's equation}
\frac{\partial^{2}(r\varphi)}{\partial r^{2}}\approx-\frac{2\pi \sigma}{\kappa}=-\frac{2\pi e}{\kappa}(n_{ext}-n).
\end{equation}
Its consequence is that the total charge within a circle of radius $r$  is approximately given by 
\begin{equation}
\label{Gauss}
\mathcal{Z}(r)\approx-\frac{\kappa}{e}r^{2}\frac{\partial\varphi}{\partial r}.
\end{equation}
Introducing a screening function $\zeta(r)$ according to
\begin{equation}
\label{screening_function}
\varphi=\frac{\zeta(r)e}{\kappa r},
\end{equation}
the charge (\ref{Gauss}) can be written as
\begin{equation}
\label{charge_within}
\mathcal{Z}(r=Re^{l})=\zeta(l)-\zeta'(l)
\end{equation}
where $R$ is an arbitrary length scale.  We see that for $l\gg 1$ ($r$ large) when the first derivative term $\zeta'(l)$ is negligible compared to $\zeta(l)$, the distinction between the screening function and the charge disappears, i.e. $\mathcal{Z}\approx\zeta$. 

\section{Thomas-Fermi theory}

In order to understand what determines the strength of screening in graphene and to go beyond the limiting cases (\ref{zero_screening_limit}) and (\ref{infinite_screening_limit}), classical electrostatics accumulated in Eqs.(\ref{sc_potential})-(\ref{n_ext_Fourier_solution}) has to be supplemented by relevant information about the physics of pristine graphene.  In the TF theory this is contained in the statement that $e\varphi$ causes a local change in the chemical potential such as in thermodynamic equilibrium the electrochemical potential remains zero across the system (the effect of finite doping has been considered in Ref.\cite{{Katsnelson}})
\begin{equation}
\label{equilibrium}
\mu=\mu_{0}(n)-e\varphi(\textbf{r})=\hbar v_{F}\sqrt{\pi n}-e\varphi(\textbf{r})=0.
\end{equation}
Here $\mu_{0}(n)=\hbar v_{F}\sqrt{\pi n}$ is the chemical potential of graphene's electrons in the absence of a perturbing potential $\varphi(\textbf{r})$ \cite{graphene_review}, and the density $n$ is assumed to be small enough so that the dispersion law (\ref{graphene_dispersion}) applies.  Eq.(\ref{equilibrium}) implies a relationship between the density $n$ and total potential $\varphi$
\begin{equation}
\label{n_of_phi_}
n=\frac{1}{\pi}\left (\frac{e\varphi}{\hbar v_{F}}\right )^{2}.
\end{equation}
For the external potential of charge $Z$ one expects either partial or complete screening, i.e. $\varphi(r\rightarrow \infty)=\mathcal{Z}(\infty)e/\kappa r$ where $0\leqslant \mathcal{Z}(\infty)< Z$ is observable charge at large distance from the source.  Taking the $r\rightarrow \infty$ limit in the first representation in Eq.(\ref{sc_potential}) one naturally finds $\mathcal{Z}(\infty)=Z-\int n(\textbf{r})d^{2}r$ which is consistent with Eq.(\ref{n_of_phi_}) only if $\mathcal{Z}(\infty)=0$.  Otherwise, the density, according to Eq.(\ref{n_of_phi_}), falls off as $1/r^{2}$ which is not normalizable.  Therefore the screening is complete, $\int n(\textbf{r})d^{2}r=Z$, with the potential and density falling off at $r$ large faster than $1/r$ and $1/r^{2}$, respectively.   

Combing Eq.(\ref{n_of_phi_}) with the first representation for $\varphi$ in Eq.(\ref{sc_potential}) leads to the nonlinear integral TF equation \cite{Katsnelson,Fogler}
\begin{equation}
\label{TF_equation_potential}
\varphi(\textbf{r})=\varphi_{ext}(\textbf{r})-\frac{e}{\pi \kappa}\left (\frac{e}{\hbar v_{F}}\right )^{2}\int \frac{\varphi^{2}(\textbf{r}')d^{2}r'}{|\textbf{r}-\textbf{r}'|}
\end{equation}
An equivalent equation can be given in terms of the density by substituting (\ref{n_of_phi_}) into the second representation in Eq.(\ref{sc_potential}) and employing the relationship (\ref{ext_potential}): 
\begin{equation}
\label{TF_equation_density}
\sqrt{n(\textbf{r})}=\frac{\alpha}{\sqrt{\pi}}\int \frac{[n_{ext}(\textbf{r}')-n(\textbf{r}')]d^{2}r'}{|\textbf{r}-\textbf{r}'|}
\end{equation}
The same equation can be obtained by minimization of the TF energy functional
\begin{eqnarray}
\label{TF_energy_functional}
E[n]&=&\frac{2\sqrt{\pi}}{3}\hbar v_{F}\int n^{3/2}(\textbf{r})d^{2}r
-e\int \varphi_{ext}(\textbf{r})n(\textbf{r})d^{2}r\nonumber\\
&+&\frac{e^{2}}{2\kappa}\int\frac{n(\textbf{r})n(\textbf{r}')d^{2}rd^{2}r'}{|\textbf{r}-\textbf{r}'|}
\end{eqnarray}  
where the first term represents quantum-mechanical effect of the zero-point motion - it is the kinetic energy of the screening cloud - while the rest encompasses classical electrostatics.  

Below we assume that the external potential is circularly-symmetric and due to charge $Z$ localized within a region of typical size $a$:
\begin{equation}
\label{smeared_external_potential}
\varphi_{ext}(\textbf{r})=\frac{Ze}{\kappa a}f_{ext}\left (\frac{r}{a}\right ),~~~f_{ext}(x\rightarrow \infty)= \frac{1}{x}
\end{equation}
where $f_{ext}(x=0)$ is finite. Measuring the length in units of $a$, potential in units of $Ze/\kappa a$ and density in units of $Z/a^{2}$ prompts introduction of dimensionless counterparts of the potential ($\varphi\rightarrow f$), density ($n\rightarrow \nu$), and external density ($n_{ext}\rightarrow \nu_{ext}$);  the latter two are now normalized at unity.  In terms of these new functions and scales Eqs.(\ref{n_of_phi_})-(\ref{TF_equation_density}) acquire the form
\begin{equation}
\label{n_of_phi_dimensionless}
\nu=\gamma f^{2},~~~\gamma=\frac{Z\alpha^{2}}{\pi}
\end{equation}
\begin{equation}
\label{dimensionless_TF_equation_potential}
f(x)=f_{ext}(x)-\gamma\int \frac{f^{2}(x')d^{2}x'}{|\textbf{x}-\textbf{x}'|}
\end{equation}
\begin{equation}
\label{dimensionless_TF_equation_density}
\sqrt{\nu(x)}=\sqrt{\gamma}\int \frac{[\nu_{ext}(x')-\nu(x')]d^{2}x'}{|\textbf{x}-\textbf{x}'|}
\end{equation}
The solution of the problem is thus determined by the shape of the external charge distribution and single dimensionless parameter $\gamma \simeq Z\alpha^{2}$ which controls the strength of screening \cite{Fogler}.  Indeed taking the infinitesimally weak screening limit $\gamma\rightarrow 0$ in Eqs.(\ref{n_of_phi_dimensionless}) and (\ref{dimensionless_TF_equation_potential}) recovers Eq.(\ref{zero_screening_limit}) while taking the infinitely strong screening limit $\gamma\rightarrow \infty$ in Eqs.(\ref{n_of_phi_dimensionless}) and (\ref{dimensionless_TF_equation_density}) derives Eq.(\ref{infinite_screening_limit}).    

It is also useful to rewrite the energy functional (\ref{TF_energy_functional}) in terms of the dimensionless density $\nu(x)$ and the external potential $f_{ext}(x)$
\begin{eqnarray}
\label{dimensionless_graphene_functional}
E[\nu]&=&\frac{Z^{2}e^{2}}{\kappa a}\Bigg\{ \frac{2}{3\sqrt{\gamma}}\int \nu^{3/2}(x)d^{2}x-\int f_{ext}(x)\nu(x)d^{2}x\nonumber\\
 &+&\frac{1}{2}\int \frac{\nu(x)\nu(x')d^{2}x d^{2}x'}{|\textbf{x}-\textbf{x}'|}\Bigg\}
\end{eqnarray}
which additionally demonstrates that the limit of infinitely strong screening $\gamma\rightarrow \infty$ is classical while quantum effects dominate interactions in the regime of infinitesimally weak screening $\gamma\rightarrow 0$.  This equation shows natural scaling of the energy with $Z$, $e$, $\kappa$ and $a$, and explains why the problem is ill-defined in the point charge limit $a=0$.  The fact that the external charge must be spread over finite-sized region imposes practical limitation on how large $Z$ can be for given $a$.  We assume that appropriate conditions discussed in Ref.\cite{Fogler} are met.

In the circularly-symmetric case of interest (\ref{smeared_external_potential}) the integral equation (\ref{TF_equation_potential}) transforms into  
\begin{eqnarray}
\label{TF_equation_circular_symmetry}
\varphi(r)&=&\varphi_{ext}(r)\nonumber\\
&-&\frac{4e}{\pi \kappa r} \left (\frac{e}{\hbar v_{F}}\right )^{2}\int_{0}^{r}\varphi^{2}(r')r'dr'\textbf{K}\left (\frac{r'}{r}\right )\nonumber\\
&-&\frac{4e}{\pi \kappa} \left (\frac{e}{\hbar v_{F}}\right )^{2}\int_{r}^{\infty}\varphi^{2}(r')dr'\textbf{K}\left (\frac{r}{r'}\right )
\end{eqnarray}
and corresponding versions of Eqs.(\ref{TF_equation_density}), (\ref{TF_energy_functional}), and (\ref{dimensionless_TF_equation_potential})-(\ref{dimensionless_graphene_functional}) can be similarly given.

In assessing the range of applicability of the TF theory we first observe that it overlooks the atomic collapse effect discussed in the Introduction.   Indeed, according to Eq.(\ref{n_of_phi_}) arbitrarily small potential induces space charge (because the energy gap is zero) while from the single-particle standpoint there is a screening threshold at $Z_{c}=1/2\alpha$.  The resolution \cite{Shytov,Fogler}, like in the problem of impurity screening in Weyl semimetal \cite{KSZ}, is that the prediction of complete screening $\mathcal{Z}(\infty)=0$, the exact property of the TF theory, is an artifact;  the observable charge for $Z\geqslant Z_{c}=1/2\alpha$ \textit{is} the critical charge $\mathcal{Z}(\infty)=1/2\alpha$.  This means that our analysis holds provided $\mathcal{Z}(r)\gg1/\alpha$ which, for $\alpha\ll1$, is more constraining than the usual semiclassical requirement of slow variation of the de Broglie wave length with position, $|dn^{-1/2}/dr| \ll 1$ \cite{LL3}.    

\subsection{Potential and density profiles in large distance limit}
 
Even though Eqs.(\ref{dimensionless_TF_equation_potential}) and (\ref{dimensionless_TF_equation_density}) appear to be good starting points of analysis in the weak, $Z\alpha^{2}\ll 1$, and strong, $Z\alpha^{2}\gg1$, screening limits, respectively, below we are going to find out that such treatments inevitably break down at sufficiently large distance from the source center.  So we begin with analysis of this large-distance regime by employing the differential equation (\ref{2dPoisson's equation}) approximating full integral equations such as (\ref{TF_equation_potential}) or (\ref{TF_equation_density}).  Combining Eq.(\ref{2dPoisson's equation}) (with $n_{ext}$ neglected) and Eq.(\ref{n_of_phi_}), and employing the screening function $\zeta(r)$ introduced in Eq.(\ref{screening_function}) we arrive at the nonlinear differential equation
\begin{equation}
\label{screening_function_equation}
\zeta''(l)-\zeta'(l)= 2\alpha^{2}\zeta^{2}
\end{equation}
whose immediate consequence is that $\zeta\propto \mathcal{Z}\propto 1/\alpha^{2}$. For $l=\ln(r/R)\ll1$ ($R$ is now a scale beyond which $n_{ext}$ in Eq.(\ref{2dPoisson's equation}) can be omitted compared to $n$), we can neglect here the first-order derivative term $\zeta'(l)$ compared to $\zeta''(l)$.   The solution to (\ref{screening_function_equation}) in this limit as well as the expression for the net charge within a circle of radius $r=Re^{l}$, $\mathcal{Z}(l)=-\zeta'(l)$ (see Eq.(\ref{charge_within}) are given by 
\begin{equation}
\label{MVP_limit_screening_function}
\zeta(l)= \frac{3}{\alpha^{2}(l+B)^{2}},~\mathcal{Z}(l)=\frac{6}{\alpha^{2}(l+B)^{3}},~0\leqslant l\ll1.
\end{equation}   
Even though Eq.(\ref{screening_function_equation}) does not mention the source charge distribution, its imprints may be present in the length scale $R$ and value of the numerical constant $B$ determined by matching (\ref{MVP_limit_screening_function}) to its $r\lesssim R$ counterpart (see below).

For $l = \ln(r/R)\gg1$, we can neglect in Eq. (\ref{screening_function_equation}) the second-order derivative term $\zeta''(l)$ compared to $\zeta'(l)$.  In this limit we do not distinguish between the charge $\mathcal{Z}$ and screening function $\zeta$ and can write down Eq.(\ref{screening_function_equation}) as the equation for the charge
\begin{equation}
\label{zero_charge_limit}
\mathcal{Z}'(l)=-2\alpha^{2}\mathcal{Z}^{2}(l)
\end{equation}
Its solution
\begin{equation}
\label{zero_charge_solution}
\mathcal{Z}=\zeta=\frac{1}{2\alpha^{2}l}=\frac{1}{2\alpha^{2}\ln(r/R)}, 1\ll \ln\frac{r}{R}\ll \frac{1}{\alpha}
\end{equation}
(with $R$ identified with a lattice length scale) was given previously \cite{Katsnelson} while the range of applicability (the charge $\mathcal{Z}$ should significantly exceed $1/\alpha$) imposes the constraint $\alpha \ll 1$ \cite{Fogler}.  Suspended graphene, $\alpha\approx 1/2$ \cite{suspended}, is at best at the verge of applicability of the asymptotic result;  the range in (\ref{zero_charge_solution}) can be made wider by choosing smaller $\alpha$ which can be accomplished by selecting large $\kappa$ substrates \cite{Fogler}.  The potential $\varphi(r)=\zeta (r)e/\kappa r$ and density (\ref{n_of_phi_}) in this regime are given by 
\begin{equation}
\label{potential_universal_limit}
\varphi(r)=\frac{e}{2\alpha^{2}\kappa r\ln(r/R)}
\end{equation}   
\begin{equation}
\label{density_universal_limit}
n(r)=\frac{1}{4\pi \alpha^{2}r^{2}\ln^{2}(r/R)}
\end{equation}
The remarkable property of Eqs.(\ref{zero_charge_solution})-(\ref{density_universal_limit}) is their near universality:  the dependence on external charge distribution is logarithmically weak and hidden in the length scale $R$ yet to be identified. 

\subsection{Weak screening:  $Z\alpha^{2}\ll 1$}
 
The condition of weak screening is consistent with the source charge being supercritical $Z\alpha \gg 1$ provided $1/\alpha \ll Z \ll 1/\alpha^{2}$ which almost certainly leaves suspended graphene, $\alpha\approx1/2$ \cite{suspended}, outside of this regime \cite{Fogler}.  For graphene on a high-$\kappa$ substrate ($\alpha \ll 1$) the weak-screening regime can be realized and the starting point of the analysis is the zero-screening limit result (\ref{zero_screening_limit}).  Then substitution of $\varphi=\varphi_{ext}$ into Eq.(\ref{n_of_phi_}) determines the electron density as the effect of the order $Z\alpha^{2}$ (see Eq.(\ref{n_of_phi_dimensionless})): 
\begin{equation}
\label{density_weak_screening}
n(r)=\frac{1}{\pi}\left (\frac{e\varphi_{ext}(r)}{\hbar v_{F}}\right )^{2}\rightarrow \frac{Z^{2}\alpha^{2}}{\pi r^{2}}
\end{equation}    
where in the second step we specified to the $r\gg a$ limit.  Inside the source charge distribution the density can be estimated as $n(r\lesssim a)\simeq Z^{2}\alpha^{2}/a^{2}$, and the number of screening electrons is of the order $(Z^{2}\alpha^{2}/a^{2})a^{2}=Z\cdot(Z\alpha^{2})\ll Z$.  We thus conclude that most of the screening charge is outside the source charge distribution.  For the total charge within a circle of radius $r\gg a$ we find with logarithmic accuracy
\begin{equation}
\label{weak_screening_charge}
\mathcal{Z}(r)=Z-2\pi \int_{0}^{r}n(r')r'dr'=Z\left (1-2Z\alpha^{2}\ln\frac{r}{a}\right )
\end{equation}
which applies provided $Z\alpha^{2}\ln(r/a)\ll 1$, i.e. it inevitably fails at sufficiently large distance from the source (the density (\ref{density_weak_screening}) is not normalizable).  

Eqs.(\ref{weak_screening_charge}) and (\ref{zero_charge_solution}) (with $R\rightarrow a$) can be combined into a compact interpolation formula that captures both perturbative short-distance and nearly-universal large-distance limits, respectively:
\begin{equation}
\label{weak_screening_interpolation_charge}
\mathcal{Z}(r)\simeq\frac{Z}{1+2Z\alpha^{2}\ln(r/a)},1\lesssim \ln\frac{r}{a}\ll \frac{1}{\alpha},Z\alpha^{2}\ll 1
\end{equation}
(and ignores the intermediate asymptotic regime (\ref{MVP_limit_screening_function})).  Same accuracy interpolation formula can be given for the electron density
\begin{equation}
\label{weak_screening_interpolation_density}
n(r)\simeq\frac{Z^{2}\alpha^{2}}{\pi r^{2}[1+2Z\alpha^{2}\ln(r/a)]^{2}}
\end{equation}
that combines Eqs.(\ref{density_universal_limit}) and (\ref{density_weak_screening}).  It is straightforward to verify that (with  accuracy controlled by the $Z\alpha^{2}\ll 1$ condition) the density distribution (\ref{weak_screening_interpolation_density}) is normalized at $Z$, and that the net charge within a circle of radius $r\gtrsim a$ computed with the help of Eq.(\ref{weak_screening_interpolation_density}) reproduces Eq.(\ref{weak_screening_interpolation_charge}).

Finally, interpolation formula for the potential
\begin{equation}
\label{weak_screening_interpolation_potential}
\varphi(r)\simeq\frac{Ze}{\kappa r[1+2Z\alpha^{2}\ln(r/a)]}
\end{equation}  
follows from the relationship (\ref{n_of_phi_}).  While Eqs. (\ref{weak_screening_interpolation_charge}), (\ref{weak_screening_interpolation_density}) and (\ref{weak_screening_interpolation_potential}) were given previously \cite{Katsnelson}, their status as interpolation formulas was not clarified;  their range of applicability as given in Eq.(\ref{weak_screening_interpolation_charge}) was pointed out in Ref.\cite{Fogler}.

\subsection{Strong screening: $Z\alpha^{2}\gg 1$}

The starting point of the analysis is the classical result (\ref{infinite_screening_limit}).  Then substitution of $n=n_{ext}$ into Eq.(\ref{n_of_phi_}) determines the potential as the effect of the order $(Z\alpha^{2})^{-1/2}$ (see Eq.(\ref{n_of_phi_dimensionless})).  This is an improvement on Eq.(\ref{infinite_screening_limit}) to the case of large finite screening parameter $Z\alpha^{2}$:
\begin{equation}
\label{large_screening_limit}
n(r)=n_{ext}(r),~~\varphi(r)=\frac{\hbar v_{F}\sqrt{\pi n_{ext}(r)}}{e},~~~~r\ll R
\end{equation}
The length scale $R$ limiting the range of applicability can be determined by recalling that in the strong screening limit the kinetic energy of the confined electrons having quantum-mechanical origin is negligible compared to the classical potential energy of the Coulomb repulsion.  Comparing the first and third term of the energy functional (\ref{TF_energy_functional}) it is straightforward to realize that they have the same order of magnitude at the scale $R$ to be self-consistently determined from the condition
\begin{equation}
\label{length_scale}
R\simeq \frac{1}{\alpha\sqrt{n_{ext}(R)}}.
\end{equation}
We note the right-hand side has the same order of magnitude as the screening length \cite{AFS} of the two-dimensional electron gas $(\kappa/2\pi e^{2})\partial \mu_{0}/\partial n$ of density $n_{ext}$ which in graphene's case is $1/4\alpha \sqrt{\pi n_{ext}}$.  It is a consequence of dimensional analysis applied to Eq.(\ref{length_scale}) that the electron charge outside the $r\simeq R$ circle can be estimated as $n(R)R^{2}\simeq 1/\alpha^{2}$ which is also an estimate for the \textit{net} charge inside the $r\simeq R$ circle, i.e. $\mathcal{Z}(R)\simeq1/\alpha^{2}$ (total electric charge of the source plus that of the electron cloud is zero in the TF theory).  Earlier (Section IIIA) $R$ was defined as a length scale beyond which $n_{ext}$ can be neglected compared to $n$.  However in that regime the charge was also found to be proportional to $1/\alpha^{2}$.   Specifically $l=0$ in the equation for the charge (\ref{MVP_limit_screening_function}) means $r=R$ thus implying that $\mathcal{Z}(R)\simeq 1/\alpha^{2}B^{3}$.  It then seems plausible that at $r\simeq R$ determined by Eq.(\ref{length_scale}), the density and potential given by Eqs.(\ref{large_screening_limit}) cross over to their $r\gtrsim R$ counterparts  as described in Section IIIA, and that the coefficient $B$ in Eq.(\ref{MVP_limit_screening_function}) is a $Z$-independent constant.  

As evidenced by Eq.(\ref{dimensionless_TF_equation_density}) relative correction to $n=n_{ext}$ has the order of $(Z\alpha^{2})^{-1/2}$ and thus is small for $r\ll R$.  Explicit calculation for the charge off the plane geometry \cite{Fogler} illustrates this assessment.

For suspended graphene, $\alpha\approx 1/2$ \cite{suspended}, $Z\gg1$, and sufficiently smooth $n_{ext}$ properties of the bulk of the electron screening cloud, $r\lesssim R$, are described by Eqs.(\ref{large_screening_limit}) and (\ref{length_scale}).  For the special case of the external point charge positioned off the graphene plane this conclusion was reached in Ref. \cite{Fogler}.  Properties of the cloud fringe, $r\gtrsim R$, may be captured semi-quantitatively by the theory of Section IIIA.    

\subsubsection{External point charge off graphene plane}

For the special case of point external charge $Z$ positioned a distance $a$ away the graphene plane the results (\ref{large_screening_limit}) and (\ref{length_scale}) were derived earlier \cite{Fogler}.  The length scale (\ref{length_scale}) in this case can be estimated as $R\simeq a\cdot(Z\alpha^{2}) \gg a$ and direct integration of the density (\ref{n_ext_charge_above_plane}) verifies our dimensional argument that the net charge within a circle of radius $R$, $\mathcal{Z}(R)=Z-2\pi \int_{0}^{R}n_{ext}(r)rdr$, is indeed of the order $1/\alpha^{2}$.  It was stated \cite{Fogler} that the conclusions 
\begin{equation}
\label{FNS_results}
n(r)=\frac{Za}{2\pi r^{3}},\varphi(r)=\frac{\hbar v_{F}}{e}\sqrt{\frac{Za}{2r^{3}}}, a\ll r\ll a\cdot(Z\alpha^{2})
\end{equation}  
originally derived for the charge off the plane geometry, specifically the power-law falloffs, $n(r) \propto 1/r^{3}$ and $\varphi(r)\propto 1/r^{3/2}$ within specified range of distances (\ref{FNS_results}), are universally applicable for generic external charge distribution of net charge $Z$ spread over a length scale $a$.  We disagree and give two experimentally relevant examples of external in-plane charge distributions.  

\subsubsection{Gaussian external charge distribution}

If the external charge distribution is Gaussian, Eqs.(\ref{n_ext_Gaussian}), (\ref{large_screening_limit}) and (\ref{length_scale}) now predict that
\begin{eqnarray}
\label{Gaussian_results}
n(r)&=&\frac{Z}{4\pi a^{2}}e^{-r^{2}/4a^{2}},\nonumber\\
\varphi(r)&=&\frac{\hbar v_{F}\sqrt{Z}}{2ea}e^{-r^{2}/8a^{2}}, r\ll R \simeq 2a\ln^{1/2}(Z\alpha^{2}).
\end{eqnarray}
We see that for Gaussian distribution of external charge the potential is also Gaussian and drastically different from the external potential (\ref{phi_ext_4_Gaussian_n_ext}).   The expression for the length scale $R$ has logarithmic accuracy and exhibits weak dependence on the screening parameter $Z\alpha^{2}$.  In practical terms the latter means that the potential is present where the external charge is.    

\subsubsection{Uniformly charged disk:  realization of two-dimensional Thomson atom}

Vacancy clusters in graphene, if created by the technique of Ref. \cite{Eva_Andrei}, would have atomically sharp boundaries.  The relevant external charge distribution is then that of an in-plane uniformly charged disk of radius $a$.  Then within the cluster the electron density and potential are constant and given by Eqs.(\ref{large_screening_limit}) (the state of local neutrality) and zero otherwise.  This solution predicts singularity of the in-plane electric field at the boundary which is removed for large but finite screening parameter $Z\alpha^{2}$.  Indeed, quantum mechanics (finite $Z\alpha^{2}$) promotes delocalization and as a result some fraction of the screening electrons moves outside of the geometrical boundary of the cluster while the majority remains inside.  As in analogous problem of impurity screening in Weyl semimetal \cite{KSZ}, the concept of the screening length (\ref{length_scale}) (with constant $n_{ext}$) continues to play an important role and gives an estimate of the width of the charge depleted region within the cluster.  Only within this layer is the external charge not locally neutralized;  small fraction of screening electrons resides outside the cluster.  This physical picture is internally consistent if the size of the depleted region $a/\sqrt{Z\alpha^{2}}$ is significantly smaller than the cluster size $a$, which indeed holds in the strong-screening limit $Z\alpha^{2}\gg 1$. 

This argument allows us to estimate the net charge within the cluster as the external charge within the electron depleted boundary region:
\begin{equation}
\label{net_charge_within_cluster}
\mathcal{Z}(a)\simeq a (Z\alpha^{2})^{-1/2}\cdot a\cdot n_{ext}\simeq Z\cdot (Z\alpha^{2})^{-1/2} \ll Z
\end{equation}
which is simultaneously the estimate for the number of screening electrons outside the cluster.  We see that majority of the screening electrons is inside the cluster while only a small fraction is outside where the description of the Section IIIA (with $R\simeq a$) applies.  Matching the charges given by Eqs.(\ref{net_charge_within_cluster}) and (\ref{MVP_limit_screening_function}) at the cluster boundary the coefficient $B$ entering Eqs.(\ref{MVP_limit_screening_function}) can be estimated as $B\simeq (Z\alpha^{2})^{-1/2}$.

Uniformly charged cluster along with screening cloud of electrons is in itself an interesting object because it provides two-dimensional realization of the Thomson atom, now obsolete model of real atom.  In our case, the "nucleus" - the bulk of the cluster - is nearly perfectly neutralized by the electrons with the exception of positively charged boundary layer while the small fraction of the screening cloud is present outside the cluster.  The boundary region forms two-dimensional electrostatic double layer where the outward-pointing electric field is localized.  The near neutrality of the bulk of the cluster is the classical effect while the double-layer structure of the boundary region has quantum-mechanical origin.  

\section{Conclusions}

Piecing together the results of analysis of various limiting cases, we arrive at the following physical picture of screening of a highly charged, $Z\alpha\gg 1$, impurity cluster in graphene:

(i) The outcome is largely dictated by the dimensionless screening parameter $Z\alpha^{2}$ and, to lesser extent, by the smoothness of the external charge distribution.  If $\alpha \ll 1$ (graphene on high-$\kappa$ substrate) there are two well-defined regimes of parameter values.

(ii) In the weak-screening case, $Z\alpha^{2} \ll 1$, most of the screening charge resides outside the cluster;  it is nearly certain that this regime cannot be realized in suspended graphene, $\alpha\approx 1/2$.  

(iii)  In the strong-screening regime, $Z\alpha^{2}\gg 1$, most of the screening charge resides inside the cluster, and this regime does apply to the case of suspended graphene.  Experimentally most relevant situation of uniformly charged cluster with sharp boundary provides two-dimensional realization of the Thomson atom.  As the screening parameter $Z\alpha^{2}$ is tuned \cite{Eva_Andrei} from small to large values, the screening cloud gradually moves inside the cluster interpolating between the regimes of weak and strong screening.     

(iv)  If $\alpha \ll 1$ the fringe of the screening cloud, regardless of the value of $Z\alpha^{2}$, exhibits nearly-universal properties.  For suspended graphene, $\alpha \approx 1/2$, this regime most likely does not exist.   

We hope that our results will motivate future experimental efforts to extend the approach of Ref.\cite{Eva_Andrei} to vacancy clusters in graphene.  These clusters could be gradually charged so that with appropriately chosen substrates the dimensionless screening parameter $Z\alpha^{2}$ could be tuned from the weak, $Z\alpha^{2}\ll 1$, to the strong, $Z\alpha^{2}\gg 1$, screening regimes.  Graphene's screening response will be monitored, including realization of a two-dimensional Thomson atom in the strong screening regime.    

\section{Acknowledgements} 

We are grateful to B. I. Shklovskii for his interest in our work.

\end{document}